\documentstyle[aps,preprint,pictex,graphicx]{revtex}

\begin{document}

\draft
\tighten

\preprint{KAIST-TH 99/02, SNUTP-99-9,hep-ph/9902291}
\title{Cosmological Gravitino Production in Gauge Mediated Supersymmetry Breaking Models}
\author{Kiwoon Choi, Kyuwan Hwang, Hang Bae Kim}
\address{Department of Physics,
	Korea Advanced Institute of Science and Technology\\
	Taejon 305-701, Korea}
\author{Taekoon Lee}
\address{Center for Theoretical Physics, Seoul National University\\
	Seoul 151-742, Korea}
\maketitle
\begin{abstract}
We study the cosmological gravitino production in gauge mediated
supersymmetry breaking models, while properly taking into
account the existence of the messenger mass scale.
It is found that for sizable parameter range of the model
the messenger sector contribution
leads to more stringent upper bound on the reheat temperature 
obtained from the condition that
the universe should not be overclosed by  relic gravitinos.
However it turns out that in the limit of relatively
low messenger scale and large gravitino mass,
the relic gravitino mass density  can be smaller than the critical density 
{\it independently of} the reheat temperature,
suggesting the possibility in this limit to have a high reheat temperature
without the dilution of gravitinos at late time.
\end{abstract}
%\pacs{PACS numbers: to be filled}

%]

\section{Introduction}

Supersymmetry (SUSY) is an elegant solution to the mass hierarchy problem
of the standard model \cite{Nilles}.
But it must be spontaneously broken because we do not see degenerate pairs
of particles and their superpartners.
Within the minimal supersymmetric standard model (MSSM), it seems unavoidable
to introduce explicit supersymmetry breaking through soft breaking terms.
Such soft breaking terms might be a consequence of spontaneous supersymmetry
breaking which is transmitted to the MSSM sector by certain mediating
interactions.  In this regards, there are two classes of models which
are distinguished by the nature of mediating interactions:
gravity-mediated models \cite{Nilles} and
gauge-mediated models \cite{Giudice-Rattazzi}.

Cosmological consequences of supersymmetry breaking models have been
also discussed to some extent.  One of the major issues has been
the relic abundance of the lightest supersymmetric particle (LSP). 
In gauge-mediated models, one interesting feature is that the gravitino
is very light and thus becomes the LSP.
For a light gravitino, its interaction is dominated by the longitudinal
goldstino component whose interaction strength is proportional to
$1/m_{3/2}$ and thus gets stronger as the gravitino mass $m_{3/2}$ becomes
smaller \cite{Fayet}.
Then the considerable amount of light gravitinos can be produced from the
decay or collisions of the MSSM particles, even in the case that the
gravitino number density was negligible just after the reheating phase of
inflation.  Not to overclose the universe, this imposes a strong constraint
on the reheat temperature depending upon the gravitino mass,
unless the gravitinos produced in this way were 
diluted by a late time entropy production
\cite{Moroi-Murayama-Yamaguchi,Gouvea-Moroi-Murayama}.

In the previous studies
\cite{Moroi-Murayama-Yamaguchi,Gouvea-Moroi-Murayama},
only the gravitino productions by the MSSM particles were considered
under the assumption that the local goldstino couplings to the MSSM fields
are valid up to energy scales around the reheat temperature $T_R$.
However it was observed by one of the authors (T.L.) 
that gravitinos decouple from
the MSSM fields at temperatures above the messinger scale at which  the
gravitino interaction to the MSSM particles is induced radiatively \cite{TLee}.
In this paper, we reexamine the cosmological gravitino production
while properly taking into account
the existence of the new mass scale in gauge-mediated models,
i.e. the messenger scale $M_X$, which is deeply involved in the problem.
We first note that the goldstino production due to the MSSM particles
at high temperature $T>M_X$ is suppressed by  $M_X^2/T^2$ when compared to
the naive extrapolation of the low temperature results at $T<M_X$.
However, in this high temperature range  the messenger sector contribution
is much larger than that of the MSSM particles because the messenger
particles have stronger couplings to the goldstino.
As a result, for much of the parameter range of $(M_X,m_{3/2})$, 
the messenger sector contribution
leads to more stringent upper bound on the reheat temperature 
which is obtained from the condition that
the universe should not be overclosed by  relic gravitinos.
The  results are  depicted  in Fig. 2 for several values of
the messenger and MSSM gaugino masses.

An interesting feature of the messenger sector contribution is that the
relic gravitino number density to the entropy number density ratio
($Y_{3/2}\equiv n_{3/2}(T)/s(T)$) behaves such as
$dY_{3/2}\propto dT/(m_{3/2}^2T^n)$ $(n\geq 2)$ for $M_X<T<T_R$,
indicating that the present value of $Y_{3/2}$ is mostly induced at
$T\sim M_X$, not at the highest available temperature $T\sim T_R$.
As a result, in gauge-mediated models with $M_X\lesssim T_R$,
the gravitino abundance is determined by the values of $m_{3/2}$
and $M_X$, and is  independent of  $T_R$.
This means that in the limit of relatively
low messenger scale
and large gravitino mass,
the relic gravitino mass density  can be smaller than the critical density 
{\it independently of} $T_R$.
Of course, this would be true only when other more model-dependent contributions,
e.g. those from the supersymmetry breaking sector, are negligible,
which would be the case if the typical mass scale of the supersymmetry
breaking sector is bigger than $T_R$.
The results of Eq.(\ref{Omega2}) and Fig. 2 show that
for the messenger mass $M_X=10^5\sim10^6$ GeV and
the gravitino mass $m_{3/2}\gtrsim 100$ MeV,
we can have $T_R$ 
much higher than the previously obtained bound
without assuming the dilution of gravitinos at late time.

\section{Goldstino couplings in  gauge-mediated models}

Gauge mediated supersymmetry breaking (GMSB) models consist of the
observable (MSSM) sector, the messenger sector, and also the supersymmetry
breaking sector \cite{Giudice-Rattazzi}.
Supersymmetry breaking sector breaks supersymmetry dynamically at certain
high energy scales and delivers supersymmetry breaking effects to the
messenger sector through messenger gauge interactions \cite{Dine-Nelson}
or other interactions \cite{others1,others2}.
Supersymmetry breaking in the messenger sector is finally transmitted to
the MSSM sector via the standard model gauge interactions and generates
the soft supersymmetry breaking masses of order $10^2\sim 10^3$ GeV.
After integrating out the supersymmetry breaking sector,
the effective superpotential is given by
\begin{equation}
W_{\rm eff} = \lambda X\Phi\Phi^c + \Delta W(X,Y_i),
\label{eq:superpotential}
\end{equation}
where $X$ and $Y_i$ are the standard model singlets and the nonperturbative
effects of the  supersymmetry breaking sector are encoded in $\Delta W$.
Here we consider the simplest messenger sector, consisting of single flavor
of chiral superfield $\Phi+\Phi^c$ transforming as $5+\bar5$ of $SU(5)$.
By minimizing the effective potential from $W_{\rm eff}$, we can determine
the vacuum expectation values (VEVs) $\langle X\rangle$, $\langle Y_i\rangle$
and also $\langle F_X^*\rangle=\langle\partial W_{\rm eff}/\partial X\rangle$,
$\langle F_i^*\rangle=\langle\partial W_{\rm eff}/\partial Y_i\rangle$
which measure the size of supersymmetry breaking.
The VEV $\langle X\rangle$ gives a supersymmetric mass
$\lambda\langle X\rangle\equiv M_X$ to the messenger $\Phi+\Phi^c$,
while $\langle F_X^*\rangle$ leads to the supersymmetry breaking mass
splitting in the scalar masses of $\Phi+\Phi^c$.
Supersymmetry breaking in the messenger sector is transmitted to the MSSM
particles via the standard model gauge interactions at one-loop or two-loop
levels, yielding the supersymmetry breaking mass splitting in MSSM particles:
\begin{equation}
m_\lambda\sim(\alpha/4\pi)F_X/\langle X\rangle,\qquad
m_{\tilde f}^2\sim(\alpha/4\pi)^2(F_X/\langle X\rangle)^2,
\label{softmass}
\end{equation}
where $m_{\lambda}$ and $m_{\tilde f}$ denote the soft gaugino mass and
scalar mass, respectively, and $\alpha$ is the MSSM fine structure constant.

The spontaneous breakdown of supersymmetry leads to a massless spin-1/2
fermions, the goldstino.  The goldstino field is a linear combination of
the fermion components of $X$ and $Y_i$ weighted by $F_X$ and $F_i$, i.e., 
\begin{equation}
\chi=\frac{\sum_I F_I^*\psi_I}{F},
\end{equation}
where $F=\sqrt{|F_X|^2+\sum_i|F_i|^2}$.
When global supersymmetry is promoted to local supersymmetry (supergravity),
the goldstino is absorbed into the gravitino as the longitudinal components 
and the gravitino acquires a mass
\begin{equation}
m_{3/2}=\frac{F}{\sqrt{3}M_P},
\end{equation}
with the vanishing cosmological constant assumed.

At energy scales below $\sqrt{F}\sim \sqrt{m_{3/2}M_P}$
but above the messenger mass scale $M_X=\lambda\langle X\rangle$,
the goldstino $\chi$  directly couples to the messenger fields,
which can be read off from the superpotential (\ref{eq:superpotential})
using the relation $\psi_I=(F_I/F)\chi+\cdots$,
\begin{eqnarray}
{\cal L}_{\rm messenger} &=& \lambda\frac{F_X}{F}
\left(\phi\chi\psi^c+\phi^c\chi\psi\right) + \cdots
\nonumber \\
&\sim&\frac{4\pi}{\alpha}\frac{m_{\lambda}M_X}{\sqrt{3}m_{3/2}M_P}
\left(\phi\chi\psi^c+\phi^c\chi\psi\right) + \cdots,
\label{messengerlagrangian}
\end{eqnarray}
where $\phi+\phi^c$ and $\psi+\psi^c$ denote the scalar and fermion components
of the messenger superfields $\Phi+\Phi^c$.
At these energy scales,
there is no direct local couplings of the goldstino to the MSSM particles
at lagrangian level.
However at lower energy scales below $M_X$, one can show \cite{TLee} that
integrating out the messenger particles through one and two loop diagrams
 leads to
the following effective lagrangian
\cite{Fayet,Lee-Wu,Brignole-Ferruccio-Zwirner}:
\begin{equation}
{\cal L}_{MSSM} =
\frac{m_{\tilde f}^2-m_f^2}{\sqrt 3m_{3/2}M_P}\bar{f}\chi\tilde{f}^*
+\frac{m_\lambda}{4\sqrt{6}m_{3/2}M_P}\bar\chi\sigma_{\mu\nu}\lambda F^{\mu\nu}
-\frac{gm_\lambda}{\sqrt{6}m_{3/2}M_P}\tilde{f}^*\tilde{f}\chi\lambda
+\textrm{h.c.},
\label{eff-lagrangian}
\end{equation}
where $(\tilde{f},f)$ and $(\lambda,F_{\mu\nu})$ denote
the MSSM matter and gauge multiplets, respectively.
We stress that the above local goldstino couplings to the MSSM fields
are valid only at energy scales below $M_X$.
When computing the amplitudes with an external energy $E\geq M_X$
(but still below $\sqrt{F}$), one has to use
(\ref{messengerlagrangian}), {\em not} (\ref{eff-lagrangian}),
even when the external particles
involve only the  goldstino and MSSM particles.
Note that the effective lagrangian (\ref{eff-lagrangian})
summarizes the leading order results  of the low energy amplitudes
expanded in powers of $E/M_X$.

For the soft masses (\ref{softmass}) presumed to be of order
$10^2\sim 10^3$ GeV, goldstino couplings
% (\ref{messengerlagrangian}) and (\ref{eff-lagrangian})
are described essentially by the two unknown model-dependent parameters,
$m_{3/2}$ and $M_X$.
In the next section, we will compute the relic gravitino mass density
as a function of $m_{3/2}$, $M_X$, and also the reheat temperature $T_R$.
The size of $m_{3/2}$ in GMSB models has a wide range of model-dependence,
which  can range from ${\cal O}(1)$ eV even to ${\cal O}(10)$ GeV.
Moreover, depending on how the messenger scale is generated and also
on how  supersymmetry breaking is transmitted
from the supersymmetry breaking sector to the messenger sector,
it is possible to have $M_X$ and $\sqrt{F_X}$ much smaller than
$\sqrt{F}\sim\sqrt{m_{3/2}M_P}$.
(Note that  $F_X/M_X$ is fixed to be of order
$4\pi m_{\lambda}/\alpha\sim 5\times 10^4$ GeV.)
% One may then have  $m_{3/2}\gtrsim 10^2$ MeV
% even for a relatively low messenger scale $M_X=10^5\sim 10^6$ GeV,
This is an interesting possibility in connection
with allowing a large reheat temperature, which will be discussed
in the next section.

\section{Cosmological Gravitino production}

At first, let us estimate the temperature range
for which the gravitinos are thermalized.
For the gravitinos to be thermalized, their reaction rate is
required to be larger than the expansion rate of the universe.
The expansion rate of the universe is measured by the Hubble parameter
$H(T)=(\pi^2g_*/90)^{1/2}T^2/M_P$ for the radiation dominated universe.
For the temperature below the supersymmetry breaking scale $\sqrt{F}\sim
\sqrt{m_{3/2}M_P}$,
the reaction rate of the gravitinos is roughly given by
\begin{equation}
\Gamma(T) \sim \sum_I\langle\sigma_I\rangle n_I(T),
\end{equation}
where $\langle\sigma_I\rangle$ is the thermal averaged cross section
producing the gravitino and $n_I(T)$ is the number density.
For $T\lesssim M_X$,
the number density of the messenger particles is Boltzmann-suppressed and
then we have to consider only the contributions from the MSSM particles,
while for $T\gtrsim M_X$, the contributions from  the messenger particles
are dominant. The resulting  reaction rate is given by 
\begin{equation}
\Gamma(T) \sim
\left\{ \begin{array}{ll}
\displaystyle
\xi_2\frac{m_\lambda^2}{m_{3/2}^2M_P^2} T^3
&\qquad {\rm for}\ T\lesssim M_X,
\\
\displaystyle
\xi_4\left(\frac{4\pi}{\alpha}\right)^2
\frac{m_\lambda^2M_X^2}{m_{3/2}^2M_P^2} T
&\qquad {\rm for}\ T\gtrsim M_X,
\end{array}\right.
\end{equation}
where $\xi_2$ and $\xi_4$ are constants of order unity.
(For $\xi_2$, $\xi_4$ and the cross sections, see the discussions  below.)
Then from $\Gamma(T)>H(T)$, we obtain the following temperature range
\begin{eqnarray}
T \gtrsim \frac{40}{\xi_2}\left(\frac{m_{3/2}}{m_\lambda}\right)^2M_P
&& \hspace{5mm}\textrm{for\ } T\lesssim M_X, \nonumber\\
T \lesssim \frac{\xi_4}{40}\left(\frac{4\pi}{\alpha}\right)^2
\left(\frac{m_\lambda}{m_{3/2}}\right)^2\frac{M_X^2}{M_P}
&& \hspace{5mm}\textrm{for\ } T\gtrsim M_X,
\end{eqnarray}
for which the gravitinos are thermalized.
Once the gravitinos were thermalized, the gravitino mass density to the
critical density ratio at present is given by\cite{pagels}
\begin{equation}
\Omega_{3/2}h^2 \sim\left(\frac{m_{3/2}}{\rm keV}\right),
\label{thermalrelic}
\end{equation}
and so $m_{3/2}$  should be less than about
$1$ keV not to overclose the universe now.
(Here $h=H_0/(100\,{\rm km}/{\rm sec}/{\rm Mpc})$
for the present Hubble expansion rate $H_0$.)

The simplest way to avoid the result (\ref{thermalrelic}) 
is to assume that gravitinos had never reached to thermal
equilibrium after the inflation, which would allow
$m_{3/2}\gtrsim 1$ keV.
Even in this case,
significant amount of  gravitinos can be produced 
by the decays or collisions
that occur in the thermal bath of the early universe subsequent to
inflation\cite{ellis},
which  usually leads to an upper bound on the inflation reheat temperature.
We can calculate the number density of such gravitinos
using the Boltzmann equation
\begin{equation}
\dot n_{3/2} + 3Hn_{3/2} =
\sum_I\langle\Gamma_{(I\rightarrow\chi+\cdots)}\rangle n_I +
\sum_{I,J}\langle\sigma_{(I+J\rightarrow\chi+\cdots)}v\rangle n_In_J
+ \cdots.
\label{eq:Boltzman}
\end{equation}
Introducing $Y_{3/2}(T)\equiv n_{3/2}(T)/s(T)$ where 
$s(T)=(2\pi^2/45)g_{*s}T^3$ is the entropy density,
and changing the variable from the time to the temperature,
we can integrate the above Boltzmann equation to obtain
\begin{equation}
Y_{3/2} \equiv \left(\frac{n_{3/2}}{s}\right)_0 = \int_{T_0}^{T_R}
dT\frac{\sum_I\langle\Gamma_{(I\rightarrow\chi+\cdots)}\rangle n_I +
\sum_{I,J}\langle\sigma_{(I+J\rightarrow\chi+\cdots)}v\rangle n_In_J
+\cdots}{s(T)H(T)T}.
\label{Yformula}
\end{equation}
Here $T_R$ is the reheat temperature after the inflation
and the RD era formula $-dt/dT=1/HT$ is used
since most contributions come during the radiation dominated era.
The evaluation of the integral of (\ref{Yformula})
for given $\Gamma_{(I\rightarrow\chi+\cdots)}$ and
$\sigma_{(I+J\rightarrow\chi+\cdots)}$
will be explained in the appendix.
The gravitino mass density produced by the decays or collisions
in the thermal bath is then given by
\begin{equation}
\Omega_{3/2}h^2\sim 2.8\times10^8\;Y_{3/2}
\left(\frac{m_{3/2}}{\rm GeV}\right).
\end{equation}

The processes which dominantly produce the gravitinos depend on the
temperature of the thermal bath.  For $T\lesssim M_X$,
MSSM particles are the dominant source of light gravitinos.
{}From the effective lagrangian (\ref{eff-lagrangian}),
$\Gamma_{(x\rightarrow y+\chi)}$ and $\sigma_{(x+y\rightarrow z+\chi)}$
where $x$, $y$, $z$ are the MSSM particles can be calculated.
When summed over all possible channels, the results are
\cite{Moroi-Murayama-Yamaguchi}
\begin{eqnarray}
\sum_{x,y}\Gamma_{(x\rightarrow y+\chi)}
&=& \xi_1\frac{m_\lambda^5}{m_{3/2}^2M_P^2}, \\
\frac12\sum_{x,y,z}\sigma_{(x+y\rightarrow z+\chi)}
&=& \xi_2\frac{m_\lambda^2}{m_{3/2}^2M_P^2},
\label{cross-section1}
\end{eqnarray}
where $\xi_1$ and $\xi_2$ are of order $1$.
We then find for $m_\lambda\lesssim T_R\lesssim M_X$
\begin{equation}
\Omega_{3/2}h^2 \sim \left(\frac{m_{3/2}}{100{\rm keV}}\right)^{-1}
\left[ 0.58\xi_1\left(\frac{m_\lambda}{10^3{\rm GeV}}\right)^3
+ 2.4\xi_2\left(\frac{m_\lambda}{10^3{\rm GeV}}\right)^2
\left(\frac{T_R}{10^5{\rm GeV}}\right)\right]
\label{Omega1}
\end{equation}

For $T\gtrsim M_X$,
the effective lagrangian (\ref{eff-lagrangian}) which gives
the cross section (\ref{cross-section1}) is not valid any more.
When computed using the correct lagrangian (\ref{messengerlagrangian}),
the cross section $\sigma_{(x+y\rightarrow z+\chi)}(s)$ falls like $1/s$
for $\sqrt{s}>M_X$.
As a result, the gravitino production due to the MSSM particles at
temperature $T>M_X$ is suppressed by  $M_X^2/T^2$ when compared to
the naive extrapolation of the result obtained from
the effective lagrangian (\ref{eff-lagrangian}).
Now the gravitinos are produced dominantly from the processes involving
the external messenger particles which couple to the goldstino more strongly.
The relevant decay widths and scattering cross sections are given by
\begin{eqnarray}
\sum_{X,X^{\prime}}\Gamma_{(X\rightarrow X^{\prime}+\chi)} &=&
\xi_3
% \left|\frac{\lambda F_X}{F}\right|^2 \;
% \frac{|\lambda F_X|^2}{M_X^3},
\left(\frac{4\pi}{\alpha}\right)^4
\frac{m_\lambda^4M_X}{3m_{3/2}^2M_P^2},
\label{decay-width2}
\nonumber
\\
\sum_{x,X,X^{\prime}}\sigma_{(x+X\rightarrow X^{\prime}+\chi)} &=&
\xi_4
% \left|\frac{\lambda F_X}{F}\right|^2 \;
\left(\frac{4\pi}{\alpha}\right)^2
\frac{m_\lambda^2M_X^2}{3m_{3/2}^2M_P^2}
\nonumber\\&&\hspace{-0mm}\times
\frac{2(2s^2-3M_X^2s+M_X^4)+s(s-2M_X^2)\log(s^2/M_X^4)}{s(s-M_X^2)^2},
\label{cross-section2}
\nonumber
\\
\frac12\sum_{x,X,X^{\prime}}\sigma_{(X+X^{\prime}\rightarrow x+\chi)} &=&
\xi_4
% \left|\frac{\lambda F_X}{F}\right|^2 \;
\left(\frac{4\pi}{\alpha}\right)^2
\frac{m_\lambda^2M_X^2}{3m_{3/2}^2M_P^2}
\nonumber\\&&\hspace{-0mm}\times
\frac{2\sqrt{s(s-4M_X^2)}
+M_X^2\log\left(\frac{s-2M_X^2-\sqrt{s(s-4M_X^2)}}
{s-2M_X^2+\sqrt{s(s-4M_X^2)}}\right)}{s(s-4M_X^2)},
\label{cross-section3}
\end{eqnarray}
where $X$, $X^{\prime}$ represent the messenger particles transforming
as $5+\bar{5}$ of $SU(5)$ and
\begin{equation}
\xi_3 = \frac{1}{4\pi}, \qquad
\xi_4 = \alpha_1+3\alpha_2+8\alpha_3.
\end{equation}
It turns out that the most important contribution comes from the messenger
particles at $T\sim M_X$, and thus we need a more careful evaluation of
the production cross sections for $s={\cal O}(M_X^2)$.
Now the relic gravitino abundance from the messenger particles
can be obtained using the equations in the appendix.
For $T_R\gtrsim M_X$, we find 
\begin{equation}
\Omega_{3/2}h^2 \sim \left(\frac{m_{3/2}}{{\rm GeV}}\right)^{-1}
\left[ 42\xi_3
\left(\frac{m_\lambda}{10^3{\rm GeV}}\right)^4
\left(\frac{M_X}{10^5{\rm GeV}}\right)^{-1}
+ 1.8\xi_4
\left(\frac{m_\lambda}{10^3{\rm GeV}}\right)^2
\left(\frac{M_X}{10^5{\rm GeV}}\right)
\right].
\label{Omega2}
\end{equation}

In Fig.~1, we plotted $\Omega_{3/2}h^2$ including both the
MSSM contributions and the messenger contributions
as a function of the
reheat temperature $T_R$ for typical values of the gravitino,
messenger and MSSM gaugino masses.
It shows that 
$\Omega_{3/2}$
for $T_R\lesssim 0.1 M_X$
is from the MSSM contribution,
while $\Omega_{3/2}$ for $T_R\gtrsim 0.1 M_X$ is essentially
from the messenger sector contribution.
Note that $\Omega_{3/2}$ is almost independent of $T_R$
for $T_R\gtrsim M_X$ as summarized in
(\ref{Omega2}).
This is because $Y_{3/2}\equiv n_{3/2}(T)/s(T)$
produced by the messenger particles behaves such as
\begin{equation}
dY_{3/2} \propto \frac{dT}{m_{3/2}^2T^n} \quad (n\geq 2),
\end{equation}
and so the present value of $Y_{3/2}$ is mostly induced at $T\sim M_X$,
not at the highest available temperature $T_R$.
This feature requires a more careful treatment of the integral 
(\ref{Yformula}) whose integrand takes a nontrivial form for $T\sim M_X$. 

Now we can obtain an upper bound on $T_R$
as a function of the gravitino mass
by requiring $\Omega_{3/2}h^2<1$.
We present the results in Fig.~2
for several values  of the messenger and MSSM gaugino masses.
Summarizing the results,
for the gravitino mass range
\begin{equation}
m_{3/2}\lesssim
0.1
\left(\frac{m_\lambda}{10^3{\rm GeV}}\right)^2
\left(\frac{M_X}{10^5{\rm GeV}}\right)
\;{\rm MeV},
\end{equation}
we find the practically same result 
as the previous one obtained considering just the MSSM 
contributions\cite{Moroi-Murayama-Yamaguchi}.
For larger $m_{3/2}$, the messenger sector contribution
leads to a significantly stronger bound on the reheat temperature:
\begin{equation}
T_R \lesssim 0.1M_X
\end{equation}
for the gravitino mass range
\begin{equation}
0.1
\left(\frac{m_\lambda}{10^3{\rm GeV}}\right)^2
\left(\frac{M_X}{10^5{\rm GeV}}\right)
\;{\rm MeV}
\lesssim m_{3/2} \lesssim
10^3 \left(\frac{m_\lambda}{10^3{\rm GeV}}\right)^2
\left(\frac{M_X}{10^5{\rm GeV}}\right)
\;{\rm MeV}.
\end{equation}
However no constraint is imposed on $T_R$ 
for the gravitino mass range
\begin{equation}
m_{3/2} \gtrsim 2
\left(\frac{m_\lambda}{10^3{\rm GeV}}\right)^2
\left(\frac{M_X}{10^5{\rm GeV}}\right)
\;{\rm GeV},
\end{equation}
suggesting that in gauge mediated models with a small messenger mass
($M_X=10^5\sim 10^6$ GeV) and also a relatively large gravitino mass
($m_{3/2}\gtrsim 100$ MeV), it may be possible to have a rather high reheat
temperature without the late dilution of relic gravitinos.
Of course, this is possible only when other more model-dependent
contributions, e.g. those from the supersymmetry breaking sector,
are small enough, which would be the case if the typical mass scale
of the supersymmetry breaking sector is bigger than $T_R$.

\medskip

{\bf Acknowledgments:} This work is supported in part by KOSEF Grant
981-0201-004-2, KOSEF through CTP of Seoul National University,
KRF under the Grant 1998-015-D00071 and the Distinguished Scholar Exchange
Program.

\medskip

\section*{Appendix}

In this appendix, we present the formulae for the integral
of (\ref{Yformula}) when the decay width $\Gamma(E)$ and the cross section
$\sigma(s)$ are given.
The decay and scattering terms in the Boltzmann equation can be written as
\begin{eqnarray}
\langle\Gamma(E)\rangle n &=& \Gamma \frac{mT^2}{2\pi^2}
\int_{m/T}^\infty dx\; \frac{(x^2-m^2/T^2)^{1/2}}{e^x\mp1},
% \\ &\simeq& \left[\frac{1}{2}\right] \frac{1}{12}mT^2\Gamma
% \qquad {\rm for}\ m\ll T
\nonumber \\
\langle\sigma_{(1+2\rightarrow\cdots)}v\rangle n_In_J &=&
\frac{T^2}{16\pi^4} \int_{(m_1+m_2)/T}^\infty dx\;
K_1(x) \sigma(x^2T^2)
\nonumber \\
&&\times \left[(x^2T^2-m_1^2-m_2^2)^2-4m_1^2m_2^2\right],
\nonumber
\end{eqnarray}
where $\Gamma$ is the decay width in the rest frame
and $K_1$ is the Bessel function.
Inserting these into (\ref{Yformula})
and changing the order of integration, we obtain
\begin{eqnarray*}
Y_{\rm decay} &\equiv& \int_{T_0}^{T_R}
\frac{\langle\Gamma_{(I\rightarrow\chi+\cdots)}\rangle n_I}{s(T)H(T)T}dT
\\
&=&
\frac{\bar gM_P\Gamma}{16\pi^2m^2}
\int_{t_R}^\infty dt\; \frac{1}{e^t\mp1}
%%\\&&\times
\left[\left\{\frac{\pi}{2}
-\tan^{-1}\left(\frac{t_R}{\sqrt{t^2-t_R^2}}\right)\right\}t^4
+t_R\left(t^2-2t_R^2\right)\sqrt{t^2-t_R^2}\right],
\\
Y_{\rm scat} &\equiv& \int_{T_0}^{T_R}
\frac{\langle\sigma_{(I+J\rightarrow\chi+\cdots)}v\rangle n_In_J}{s(T)H(T)T}dT
\\
&=&
\frac{\bar gM_P}{16\pi^4} \int_{t_R}^\infty dt\; t^3K_1(t)
%%\\&&\times
\int_{(m_1+m_2)}^{tT_R} d(\sqrt{s}) \; \sigma(s)
\left[\frac{(s-m_1^2-m_2^2)^2-4m_1^2m_2^2}{s^2}\right],
\end{eqnarray*}
where $\bar g=135\sqrt{10}/(2\pi^3g_*^{3/2})$ and
$t_R=m/T_R$ in $Y_{\rm decay}$ and $(m_1+m_2)/T_R$ in $Y_{\rm Scat}$.
For $Y_{\rm scat}$, the Maxwell-Boltzmann distribution is used
for the simplicity of calculation,
which gives the correct result up to a factor of order 1.

For $T_R>m$, $Y_{\rm decay}$ is well approximated by
\[ %\begin{equation}
Y_{\rm decay} \simeq \frac{3\zeta(5)\bar gM_P\Gamma}{4\pi m^2}.
\] %\end{equation}
For $Y_{\rm scat}$, we present two typical examples.
First, when $m_1,m_2\ll T_R$ and $\sigma(s)=\sigma_0={\rm constant}$,
$Y_{\rm scat}$ is given by
\[ %\begin{equation}
Y_{\rm scat} = \frac{\bar g\sigma_0M_PT_R}{\pi^4}.
\] %\end{equation}
For the second example with $m_1=M_X$, $m_2=0$ and $\sigma(s)$
of (\ref{cross-section3}), we have
\begin{eqnarray*}
Y_{\rm scat} &=&
\frac{\bar g\xi_4m_\lambda^2M_X}{3\pi^2\alpha^2m_{3/2}^2M_P}
\int_{t_R}^\infty \!dt\, K_1(t)
%%\\&&\times
\left[\frac{248}{45}t^3-8t_Rt^2+\frac{62}{9}t_R^3
-\frac{2}{5}\frac{t_R^5}{t^2}
-\left(t_Rt^2-\frac{2}{3}t_R^3\right)
\log\left(\frac{t^4}{t_R^4}\right)\right]
\\
&\simeq&
\frac{124\bar g\xi_4m_\lambda^2M_X}{45\pi\alpha^2m_{3/2}^2M_P},
\qquad{\rm for}\ T_R\gg M_X.
\end{eqnarray*}

%%
%%  FIGURES
%%

\begin{figure}
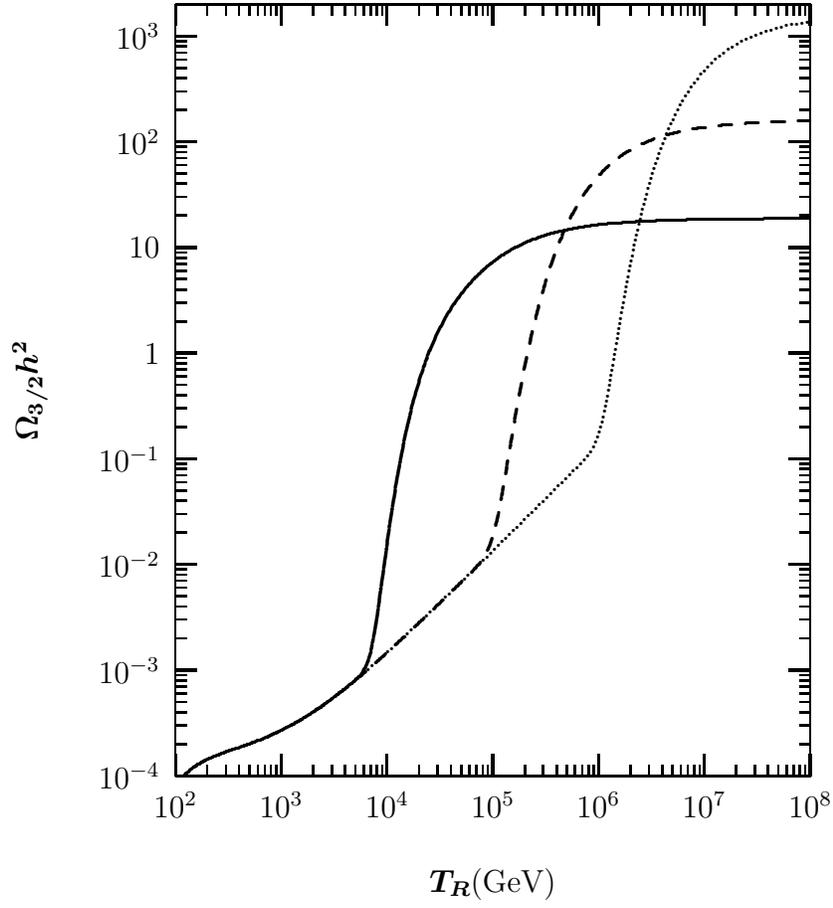

\medskip
$$\beginpicture

\setcoordinatesystem units <40pt,40pt> point at 0 0
\setplotarea  x from 2.0 to 8.0, y from -4.0 to 3.3
\inboundscheckon
\linethickness 0.5pt

%% Frame
\axis bottom label {{\boldmath$T_R$(GeV)}}
      ticks in
            width <0.8pt> length <8.0pt>
            withvalues {$10^2$} {$10^3$} {$10^4$} {$10^5$} 
                       {$10^6$} {$10^7$} {$10^8$} /
            at 2 3 4 5 6 7 8 /
            width <0.5pt> length <4.0pt> at
     2.301 2.477 2.602 2.699 2.778 2.845 2.903 2.954
     3.301 3.477 3.602 3.699 3.778 3.845 3.903 3.954
     4.301 4.477 4.602 4.699 4.778 4.845 4.903 4.954
     5.301 5.477 5.602 5.699 5.778 5.845 5.903 5.954
     6.301 6.477 6.602 6.699 6.778 6.845 6.903 6.954
     7.301 7.477 7.602 7.699 7.778 7.845 7.903 7.954
     /
/
\axis top label {}
      ticks in
            width <0.8pt> length <8.0pt>
            at 2 3 4 5 6 7 8 /
            width <0.5pt> length <4.0pt> at
     2.301 2.477 2.602 2.699 2.778 2.845 2.903 2.954
     3.301 3.477 3.602 3.699 3.778 3.845 3.903 3.954
     4.301 4.477 4.602 4.699 4.778 4.845 4.903 4.954
     5.301 5.477 5.602 5.699 5.778 5.845 5.903 5.954
     6.301 6.477 6.602 6.699 6.778 6.845 6.903 6.954
     7.301 7.477 7.602 7.699 7.778 7.845 7.903 7.954
     /
/
\axis left label {\rotatebox[origin=c]{90}{\boldmath$\Omega_{3/2} h^2$}}
      ticks in
            width <0.8pt> length <8.0pt>
            withvalues {$10^{-4}$} {$10^{-3}$} {$10^{-2}$} {$10^{-1}$}
	               {$1$} {$10$} {$10^2$} {$10^3$} /
            at -4 -3 -2 -1 0 1 2 3 /
            width <0.5pt> length <4.0pt> at
     -3.699 -3.523 -3.398 -3.301 -3.222 -3.155 -3.097 -3.046
     -2.699 -2.523 -2.398 -2.301 -2.222 -2.155 -2.097 -2.046
     -1.699 -1.523 -1.398 -1.301 -1.222 -1.155 -1.097 -1.046
     -0.699 -0.523 -0.398 -0.301 -0.222 -0.155 -0.097 -0.046
     0.301 0.477 0.602 0.699 0.778 0.845 0.903 0.954
     1.301 1.477 1.602 1.699 1.778 1.845 1.903 1.954
     2.301 2.477 2.602 2.699 2.778 2.845 2.903 2.954
     /
/
\axis right label {}
      ticks in
            width <0.8pt> length <8.0pt>
            at -4 -3 -2 -1 0 1 2 3 /
            width <0.5pt> length <4.0pt> at
     -3.699 -3.523 -3.398 -3.301 -3.222 -3.155 -3.097 -3.046
     -2.699 -2.523 -2.398 -2.301 -2.222 -2.155 -2.097 -2.046
     -1.699 -1.523 -1.398 -1.301 -1.222 -1.155 -1.097 -1.046
     -0.699 -0.523 -0.398 -0.301 -0.222 -0.155 -0.097 -0.046
     0.301 0.477 0.602 0.699 0.778 0.845 0.903 0.954
     1.301 1.477 1.602 1.699 1.778 1.845 1.903 1.954
     2.301 2.477 2.602 2.699 2.778 2.845 2.903 2.954
     /
/

%% Data Plot
\setplotsymbol (.)
\setquadratic

%% m_\lambda=300GeV, M_X=10^5GeV
\setsolid
\plot
	2        -4.08847 
	2.05        -4.02886 
	2.1        -3.97874 
	2.15        -3.93662 
	2.2        -3.90109 
	2.25        -3.87091 
	2.3        -3.84495 
	2.35        -3.82223 
	2.4        -3.80192 
	2.45        -3.78328 
	2.5        -3.76571 
	2.55        -3.74869 
	2.6        -3.73179 
	2.65        -3.71468 
	2.7        -3.69705 
	2.75        -3.67869 
	2.8        -3.6594 
	2.85        -3.63905 
	2.9        -3.61751 
	2.95        -3.59471 
	3.        -3.57059 
	3.05        -3.54511 
	3.1        -3.51825 
	3.15        -3.49 
	3.2        -3.46037 
	3.25        -3.42938 
	3.3        -3.39707 
	3.35        -3.36348 
	3.4        -3.32864 
	3.45        -3.29263 
	3.5        -3.25548 
	3.55        -3.21727 
	3.6        -3.17803 
	3.65        -3.13768 
	3.7        -3.09518 
	3.75        -3.04489 
	3.8        -2.9658 
	3.85        -2.8102 
	3.9        -2.5393 
	3.95        -2.18729 
	4.        -1.82028 
	4.05        -1.47663 
	4.1        -1.16893 
	4.15        -0.898488 
	4.2        -0.662698 
	4.25        -0.457782 
	4.3        -0.279766 
	4.35        -0.124847 
	4.4        0.0104636 
	4.45        0.129273 
	4.5        0.234278 
	4.55        0.327764 
	4.6        0.411609 
	4.65        0.487325 
	4.7        0.556098 
	4.75        0.618846 
	4.8        0.676274 
	4.85        0.72893 
	4.9        0.777248 
	4.95        0.821584 
	5.        0.862244 
	5.05        0.898685 
	5.1        0.932753 
	5.15        0.96389 
	5.2        0.992316 
	5.25        1.01824 
	5.3        1.04186 
	5.35        1.06336 
	5.4        1.08293 
	5.45        1.10071 
	5.5        1.11688 
	5.55        1.13156 
	5.6        1.1449 
	5.65        1.15701 
	5.7        1.16801 
	5.75        1.17799 
	5.8        1.18705 
	5.85        1.19527 
	5.9        1.20274 
	5.95        1.20952 
	6.        1.21567 
	6.05        1.22125 
	6.1        1.22632 
	6.15        1.23092 
	6.2        1.23509 
	6.25        1.23889 
	6.3        1.24233 
	6.35        1.24545 
	6.4        1.24828 
	6.45        1.25085 
	6.5        1.25319 
	6.55        1.25531 
	6.6        1.25723 
	6.65        1.25897 
	6.7        1.26055 
	6.75        1.26199 
	6.8        1.26329 
	6.85        1.26447 
	6.9        1.26554 
	6.95        1.26651 
	7.        1.26739 
	7.05        1.26818 
	7.1        1.2689 
	7.15        1.26956 
	7.2        1.27015 
	7.25        1.27068 
	7.3        1.27117 
	7.35        1.27161 
	7.4        1.27201 
	7.45        1.27237 
	7.5        1.27269 
	7.55        1.27299 
	7.6        1.27325 
	7.65        1.27349 
	7.7        1.27371 
	7.75        1.27391 
	7.8        1.27409 
	7.85        1.27425 
	7.9        1.27439 
	7.95        1.27453 
	8.        1.27464 
/

%% m_\lambda=300GeV, M_X=10^6GeV
\setdashes
\plot
	2        -4.08847 
	2.05        -4.02886 
	2.1        -3.97874 
	2.15        -3.93662 
	2.2        -3.90109 
	2.25        -3.87091 
	2.3        -3.84495 
	2.35        -3.82223 
	2.4        -3.80192 
	2.45        -3.78328 
	2.5        -3.76571 
	2.55        -3.74869 
	2.6        -3.73179 
	2.65        -3.71468 
	2.7        -3.69705 
	2.75        -3.67869 
	2.8        -3.6594 
	2.85        -3.63905 
	2.9        -3.61751 
	2.95        -3.59471 
	3.        -3.57059 
	3.05        -3.54511 
	3.1        -3.51825 
	3.15        -3.49 
	3.2        -3.46037 
	3.25        -3.42938 
	3.3        -3.39707 
	3.35        -3.36348 
	3.4        -3.32864 
	3.45        -3.29263 
	3.5        -3.25548 
	3.55        -3.21727 
	3.6        -3.17805 
	3.65        -3.1379 
	3.7        -3.09686 
	3.75        -3.05501 
	3.8        -3.0124 
	3.85        -2.9691 
	3.9        -2.92516 
	3.95       -2.88063 
	4.         -2.83557 
	4.05        -2.79001 
	4.1        -2.74401 
	4.15        -2.69761 
	4.2        -2.65084 
	4.25        -2.60374 
	4.3        -2.55635 
	4.35        -2.50868 
	4.4        -2.46077 
	4.45        -2.41264 
	4.5        -2.36431 
	4.55        -2.31581 
	4.6        -2.26715 
	4.65        -2.21834 
	4.7        -2.16938 
	4.75        -2.12018 
	4.8        -2.07021 
	4.85        -2.01752 
	4.9        -1.95603 
	4.95        -1.8711 
	5.        -1.73812 
	5.05        -1.53683 
	5.1        -1.2746 
	5.15        -0.981866 
	5.2        -0.687168 
	5.25        -0.406684 
	5.3        -0.147159 
	5.35        0.0895886 
	5.4         0.303963 
	5.45        0.49729 
	5.5        0.671227 
	5.55        0.827495 
	5.6        0.967767 
	5.65        1.09361 
	5.7        1.20648 
	5.75        1.30768 
	5.8        1.39843 
	5.85        1.47981 
	5.9        1.55279 
	5.95        1.61826 
	6.         1.67701 
	6.05        1.72853 
	6.1        1.77587 
	6.15        1.81839 
	6.2        1.85658 
	6.25        1.89091 
	6.3        1.92179 
	6.35        1.94957 
	6.4        1.97458 
	6.45        1.9971 
	6.5        2.0174 
	6.55        2.0357 
	6.6        2.05221 
	6.65        2.06711 
	6.7        2.08057 
	6.75        2.09272 
	6.8        2.10371 
	6.85        2.11364 
	6.9        2.12263 
	6.95        2.13076 
	7.        2.13812 
	7.05        2.14479 
	7.1        2.15082 
	7.15        2.15629 
	7.2        2.16124 
	7.25        2.16573 
	7.3        2.1698 
	7.35        2.17349 
	7.4        2.17683 
	7.45        2.17986 
	7.5        2.18261 
	7.55        2.1851 
	7.6        2.18736 
	7.65        2.1894 
	7.7        2.19126 
	7.75        2.19294 
	7.8        2.19446 
	7.85        2.19585 
	7.9        2.1971 
	7.95        2.19823 
	8.        2.19926 
/

%% m_\lambda=300GeV, M_X=10^7GeV
\setdots <2pt>
\plot
	2        -4.08847 
	2.05        -4.02886 
	2.1        -3.97874 
	2.15        -3.93662 
	2.2        -3.90109 
	2.25        -3.87091 
	2.3        -3.84495 
	2.35        -3.82223 
	2.4        -3.80192 
	2.45        -3.78328 
	2.5        -3.76571 
	2.55        -3.74869 
	2.6        -3.73179 
	2.65        -3.71468 
	2.7        -3.69705 
	2.75        -3.67869 
	2.8        -3.6594 
	2.85        -3.63905 
	2.9        -3.61751 
	2.95        -3.59471 
	3.        -3.57059 
	3.05        -3.54511 
	3.1        -3.51825 
	3.15        -3.49 
	3.2        -3.46037 
	3.25        -3.42938 
	3.3        -3.39707 
	3.35        -3.36348 
	3.4        -3.32864 
	3.45        -3.29263 
	3.5        -3.25548 
	3.55        -3.21727 
	3.6        -3.17805 
	3.65        -3.1379 
	3.7        -3.09686 
	3.75        -3.05501 
	3.8        -3.0124 
	3.85        -2.9691 
	3.9        -2.92516 
	3.95        -2.88063 
	4.        -2.83557 
	4.05        -2.79001 
	4.1        -2.74401 
	4.15        -2.69761 
	4.2        -2.65084 
	4.25        -2.60374 
	4.3        -2.55635 
	4.35        -2.50868 
	4.4        -2.46077 
	4.45        -2.41264 
	4.5        -2.36431 
	4.55        -2.31581 
	4.6        -2.26715 
	4.65        -2.21834 
	4.7        -2.16941 
	4.75        -2.12037 
	4.8        -2.07122 
	4.85        -2.02198 
	4.9        -1.97266 
	4.95        -1.92327 
	5.        -1.87381 
	5.05        -1.8243 
	5.1        -1.77473 
	5.15        -1.72511 
	5.2        -1.67545 
	5.25        -1.62576 
	5.3        -1.57603 
	5.35        -1.52628 
	5.4        -1.47649 
	5.45        -1.42669 
	5.5        -1.37686 
	5.55        -1.32701 
	5.6        -1.27715 
	5.65        -1.22727 
	5.7        -1.17737 
	5.75        -1.12741 
	5.8        -1.07711 
	5.85        -1.02525 
	5.9        -0.967523 
	5.95        -0.891785 
	6.        -0.773777 
	6.05        -0.586798 
	6.1        -0.329933 
	6.15        -0.033564 
	6.2        0.268839 
	6.25        0.55749 
	6.3        0.82419 
	6.35        1.0668 
	6.4        1.28581 
	6.45        1.48277 
	6.5        1.65952 
	6.55        1.81799 
	6.6        1.95997 
	6.65        2.08715 
	6.7        2.20107 
	6.75        2.30311 
	6.8        2.39452 
	6.85        2.47643 
	6.9        2.54984 
	6.95        2.61566 
	7.        2.67469 
	7.05        2.72644 
	7.1        2.77398 
	7.15        2.81665 
	7.2        2.85499 
	7.25        2.88943 
	7.3        2.9204 
	7.35        2.94826 
	7.4        2.97334 
	7.45        2.99593 
	7.5        3.01628 
	7.55        3.03462 
	7.6        3.05117 
	7.65        3.06611 
	7.7        3.0796 
	7.75        3.09178 
	7.8        3.10279 
	7.85        3.11274 
	7.9        3.12175 
	7.95        3.1299 
	8.        3.13727 
/

% \setdashpattern <5pt,2.4pt,0.2pt,2.4pt>

\endpicture$$
\medskip
\caption{
$\Omega_{3/2}h^2$ as a function of the reheat temperature $T_R$
for the gravitino mass $m_{3/2}=10$ MeV,
the MSSM gaugino mass $m_\lambda=300$ GeV and
the messenger mass $M_X=10^5$ GeV (solid line),
$10^6$ GeV (dashed line), $10^7$ GeV (dotted line).
$\Omega_{3/2}h^2$ is scaled by $(10 \, {\rm MeV}/m_{3/2})$ for different 
values of $m_{3/2}$.
}
\end{figure}

\begin{figure}
\medskip
$$\beginpicture

\setcoordinatesystem units <36pt,36pt> point at 0 0
\setplotarea x from -6.0 to 1.0, y from 2.0 to 8.0
\inboundscheckon
\linethickness 0.5pt

%% Frame
\axis bottom label {\boldmath$m_{3/2}$(GeV)}
      ticks in
            width <0.8pt> length <8.0pt>
            withvalues {$10^{-6}$} {$10^{-5}$} {$10^{-4}$} {$10^{-3}$} 
                       {$10^{-2}$} {$10^{-1}$} {$1$} {$10$} /
            at -6 -5 -4 -3 -2 -1 0 1 /
            width <0.5pt> length <4.0pt> at
     -5.699 -5.523 -5.398 -5.301 -5.222 -5.155 -5.097 -5.046
     -4.699 -4.523 -4.398 -4.301 -4.222 -4.155 -4.097 -4.046
     -3.699 -3.523 -3.398 -3.301 -3.222 -3.155 -3.097 -3.046
     -2.699 -2.523 -2.398 -2.301 -2.222 -2.155 -2.097 -2.046
     -1.699 -1.523 -1.398 -1.301 -1.222 -1.155 -1.097 -1.046
     -0.699 -0.523 -0.398 -0.301 -0.222 -0.155 -0.097 -0.046
     0.301 0.477 0.602 0.699 0.778 0.845 0.903 0.954
     /
/
\axis left label {\rotatebox[origin=c]{90}{\boldmath$T_R$(GeV)}}
      ticks in
            width <0.8pt> length <8.0pt>
            withvalues {$10^2$} {$10^3$} {$10^4$} {$10^5$} 
                       {$10^6$} {$10^7$} {$10^8$} /
            at 2 3 4 5 6 7 8 /
            width <0.5pt> length <4.0pt> at
     2.301 2.477 2.602 2.699 2.778 2.845 2.903 2.954
     3.301 3.477 3.602 3.699 3.778 3.845 3.903 3.954
     4.301 4.477 4.602 4.699 4.778 4.845 4.903 4.954
     5.301 5.477 5.602 5.699 5.778 5.845 5.903 5.954
     6.301 6.477 6.602 6.699 6.778 6.845 6.903 6.954
     7.301 7.477 7.602 7.699 7.778 7.845 7.903 7.954
     /
/
\axis top
      ticks in
            width <0.8pt> length <8.0pt>
            at -6 -5 -4 -3 -2 -1 0 1 /
            width <0.5pt> length <4.0pt> at
     -5.699 -5.523 -5.398 -5.301 -5.222 -5.155 -5.097 -5.046
     -4.699 -4.523 -4.398 -4.301 -4.222 -4.155 -4.097 -4.046
     -3.699 -3.523 -3.398 -3.301 -3.222 -3.155 -3.097 -3.046
     -2.699 -2.523 -2.398 -2.301 -2.222 -2.155 -2.097 -2.046
     -1.699 -1.523 -1.398 -1.301 -1.222 -1.155 -1.097 -1.046
     -0.699 -0.523 -0.398 -0.301 -0.222 -0.155 -0.097 -0.046
     0.301 0.477 0.602 0.699 0.778 0.845 0.903 0.954
     /
/
\axis right label { }
      ticks in
            width <0.8pt> length <8.0pt>
            at 2 3 4 5 6 7 8 /
            width <0.5pt> length <4.0pt> at
     2.301 2.477 2.602 2.699 2.778 2.845 2.903 2.954
     3.301 3.477 3.602 3.699 3.778 3.845 3.903 3.954
     4.301 4.477 4.602 4.699 4.778 4.845 4.903 4.954
     5.301 5.477 5.602 5.699 5.778 5.845 5.903 5.954
     6.301 6.477 6.602 6.699 6.778 6.845 6.903 6.954
     7.301 7.477 7.602 7.699 7.778 7.845 7.903 7.954
     /
/

\put {(a)} at -5 7

%% Data Plot
\setplotsymbol (.)
\setquadratic

%% m_\lambda=300GeV, M_X=10^5GeV
\setsolid
\plot
	-6.08847        2 
	-6.02886        2.05 
	-5.97874        2.1 
	-5.93662        2.15 
	-5.90109        2.2 
	-5.87091        2.25 
	-5.84495        2.3 
	-5.82223        2.35 
	-5.80192        2.4 
	-5.78328        2.45 
	-5.76571        2.5 
	-5.74869        2.55 
	-5.73179        2.6 
	-5.71468        2.65 
	-5.69705        2.7 
	-5.67869        2.75 
	-5.6594        2.8 
	-5.63905        2.85 
	-5.61751        2.9 
	-5.59471        2.95 
	-5.57059        3. 
	-5.54511        3.05 
	-5.51825        3.1 
	-5.49        3.15 
	-5.46037        3.2 
	-5.42938        3.25 
	-5.39707        3.3 
	-5.36348        3.35 
	-5.32864        3.4 
	-5.29263        3.45 
	-5.25548        3.5 
	-5.21727        3.55 
	-5.17803        3.6 
	-5.13768        3.65 
	-5.09518        3.7 
	-5.04489        3.75 
	-4.9658        3.8 
	-4.8102        3.85 
	-4.5393        3.9 
	-4.18729        3.95 
	-3.82028        4. 
	-3.47663        4.05 
	-3.16893        4.1 
	-2.89849        4.15 
	-2.6627        4.2 
	-2.45778        4.25 
	-2.27977        4.3 
	-2.12485        4.35 
	-1.98954        4.4 
	-1.87073        4.45 
	-1.76572        4.5 
	-1.67224        4.55 
	-1.58839        4.6 
	-1.51267        4.65 
	-1.4439        4.7 
	-1.38115        4.75 
	-1.32373        4.8 
	-1.27107        4.85 
	-1.22275        4.9 
	-1.17842        4.95 
	-1.13776        5. 
	-1.10132        5.05 
	-1.06725        5.1 
	-1.03611        5.15 
	-1.00768        5.2 
	-0.98176        5.25 
	-0.958139        5.3 
	-0.936636        5.35 
	-0.917073        5.4 
	-0.899287        5.45 
	-0.883122        5.5 
	-0.868437        5.55 
	-0.855099        5.6 
	-0.842989        5.65 
	-0.831993        5.7 
	-0.822012        5.75 
	-0.812951        5.8 
	-0.804726        5.85 
	-0.797261        5.9 
	-0.790484        5.95 
	-0.784333        6. 
	-0.77875        6.05 
	-0.773682        6.1 
	-0.769081        6.15 
	-0.764905        6.2 
	-0.761115        6.25 
	-0.757674        6.3 
	-0.754551        6.35 
	-0.751717        6.4 
	-0.749145        6.45 
	-0.746811        6.5 
	-0.744693        6.55 
	-0.742771        6.6 
	-0.741028        6.65 
	-0.739446        6.7 
	-0.738012        6.75 
	-0.736711        6.8 
	-0.735531        6.85 
	-0.734462        6.9 
	-0.733493        6.95 
	-0.732614        7. 
	-0.731818        7.05 
	-0.731096        7.1 
	-0.730443        7.15 
	-0.729851        7.2 
	-0.729315        7.25 
	-0.72883        7.3 
	-0.72839        7.35 
	-0.727993        7.4 
	-0.727633        7.45 
	-0.727307        7.5 
	-0.727013        7.55 
	-0.726746        7.6 
	-0.726505        7.65 
	-0.726287        7.7 
	-0.72609        7.75 
	-0.725912        7.8 
	-0.725752        7.85 
	-0.725606        7.9 
	-0.725475        7.95 
	-0.725356        8. 
/

%% m_\lambda=300GeV, M_X=10^6GeV
\setdashes
\plot
	-6.08847        2 
	-6.02886        2.05 
	-5.97874        2.1 
	-5.93662        2.15 
	-5.90109        2.2 
	-5.87091        2.25 
	-5.84495        2.3 
	-5.82223        2.35 
	-5.80192        2.4 
	-5.78328        2.45 
	-5.76571        2.5 
	-5.74869        2.55 
	-5.73179        2.6 
	-5.71468        2.65 
	-5.69705        2.7 
	-5.67869        2.75 
	-5.6594        2.8 
	-5.63905        2.85 
	-5.61751        2.9 
	-5.59471        2.95 
	-5.57059        3. 
	-5.54511        3.05 
	-5.51825        3.1 
	-5.49        3.15 
	-5.46037        3.2 
	-5.42938        3.25 
	-5.39707        3.3 
	-5.36348        3.35 
	-5.32864        3.4 
	-5.29263        3.45 
	-5.25548        3.5 
	-5.21727        3.55 
	-5.17805        3.6 
	-5.1379        3.65 
	-5.09686        3.7 
	-5.05501        3.75 
	-5.0124        3.8 
	-4.9691        3.85 
	-4.92516        3.9 
	-4.88063        3.95 
	-4.83557        4. 
	-4.79001        4.05 
	-4.74401        4.1 
	-4.69761        4.15 
	-4.65084        4.2 
	-4.60374        4.25 
	-4.55635        4.3 
	-4.50868        4.35 
	-4.46077        4.4 
	-4.41264        4.45 
	-4.36431        4.5 
	-4.31581        4.55 
	-4.26715        4.6 
	-4.21834        4.65 
	-4.16938        4.7 
	-4.12018        4.75 
	-4.07021        4.8 
	-4.01752        4.85 
	-3.95603        4.9 
	-3.8711        4.95 
	-3.73812        5. 
	-3.53683        5.05 
	-3.2746        5.1 
	-2.98187        5.15 
	-2.68717        5.2 
	-2.40668        5.25 
	-2.14716        5.3 
	-1.91041        5.35 
	-1.69604        5.4 
	-1.50271        5.45 
	-1.32877        5.5 
	-1.1725        5.55 
	-1.03223        5.6 
	-0.906389        5.65 
	-0.793524        5.7 
	-0.692318        5.75 
	-0.601568        5.8 
	-0.520191        5.85 
	-0.447207        5.9 
	-0.381738        5.95 
	-0.322993        6. 
	-0.271467        6.05 
	-0.224127        6.1 
	-0.181613        6.15 
	-0.143418        6.2 
	-0.109086        6.25 
	-0.0782115        6.3 
	-0.0504322        6.35 
	-0.0254248        6.4 
	-0.00290088        6.45 
	0.017397        6.5 
	0.0356984        6.55 
	0.0522083        6.6 
	0.0671095        6.65 
	0.0805653        6.7 
	0.0927216        6.75 
	0.103709        6.8 
	0.113643        6.85 
	0.12263        6.9 
	0.130761        6.95 
	0.138121        7. 
	0.144785        7.05 
	0.15082        7.1 
	0.156288        7.15 
	0.161241        7.2 
	0.16573        7.25 
	0.169798        7.3 
	0.173486        7.35 
	0.176829        7.4 
	0.179859        7.45 
	0.182607        7.5 
	0.185097        7.55 
	0.187355        7.6 
	0.189402        7.65 
	0.191258        7.7 
	0.19294        7.75 
	0.194465        7.8 
	0.195846        7.85 
	0.197098        7.9 
	0.198233        7.95 
	0.199261        8. 
/

%% m_\lambda=300GeV, M_X=10^7GeV
\setdots <2pt>
\plot
	-6.08847        2 
	-6.02886        2.05 
	-5.97874        2.1 
	-5.93662        2.15 
	-5.90109        2.2 
	-5.87091        2.25 
	-5.84495        2.3 
	-5.82223        2.35 
	-5.80192        2.4 
	-5.78328        2.45 
	-5.76571        2.5 
	-5.74869        2.55 
	-5.73179        2.6 
	-5.71468        2.65 
	-5.69705        2.7 
	-5.67869        2.75 
	-5.6594        2.8 
	-5.63905        2.85 
	-5.61751        2.9 
	-5.59471        2.95 
	-5.57059        3. 
	-5.54511        3.05 
	-5.51825        3.1 
	-5.49        3.15 
	-5.46037        3.2 
	-5.42938        3.25 
	-5.39707        3.3 
	-5.36348        3.35 
	-5.32864        3.4 
	-5.29263        3.45 
	-5.25548        3.5 
	-5.21727        3.55 
	-5.17805        3.6 
	-5.1379        3.65 
	-5.09686        3.7 
	-5.05501        3.75 
	-5.0124        3.8 
	-4.9691        3.85 
	-4.92516        3.9 
	-4.88063        3.95 
	-4.83557        4. 
	-4.79001        4.05 
	-4.74401        4.1 
	-4.69761        4.15 
	-4.65084        4.2 
	-4.60374        4.25 
	-4.55635        4.3 
	-4.50868        4.35 
	-4.46077        4.4 
	-4.41264        4.45 
	-4.36431        4.5 
	-4.31581        4.55 
	-4.26715        4.6 
	-4.21834        4.65 
	-4.16941        4.7 
	-4.12037        4.75 
	-4.07122        4.8 
	-4.02198        4.85 
	-3.97266        4.9 
	-3.92327        4.95 
	-3.87381        5. 
	-3.8243        5.05 
	-3.77473        5.1 
	-3.72511        5.15 
	-3.67545        5.2 
	-3.62576        5.25 
	-3.57603        5.3 
	-3.52628        5.35 
	-3.47649        5.4 
	-3.42669        5.45 
	-3.37686        5.5 
	-3.32701        5.55 
	-3.27715        5.6 
	-3.22727        5.65 
	-3.17737        5.7 
	-3.12741        5.75 
	-3.07711        5.8 
	-3.02525        5.85 
	-2.96752        5.9 
	-2.89178        5.95 
	-2.77378        6. 
	-2.5868        6.05 
	-2.32993        6.1 
	-2.03356        6.15 
	-1.73116        6.2 
	-1.44251        6.25 
	-1.17581        6.3 
	-0.933202        6.35 
	-0.714187        6.4 
	-0.517233        6.45 
	-0.340476        6.5 
	-0.182014        6.55 
	-0.0400324        6.6 
	0.0871504        6.65 
	0.201068        6.7 
	0.303108        6.75 
	0.394521        6.8 
	0.47643        6.85 
	0.549842        6.9 
	0.61566        6.95 
	0.67469        7. 
	0.726444        7.05 
	0.773979        7.1 
	0.816654        7.15 
	0.854985        7.2 
	0.889431        7.25 
	0.920403        7.3 
	0.948264        7.35 
	0.973342        7.4 
	0.995927        7.45 
	1.01628        7.5 
	1.03462        7.55 
	1.05117        7.6 
	1.06611        7.65 
	1.0796        7.7 
	1.09178        7.75 
	1.10279        7.8 
	1.11274        7.85 
	1.12175        7.9 
	1.1299        7.95 
	1.13727        8. 
/

\endpicture$$
\medskip
$$\beginpicture

\setcoordinatesystem units <36pt,36pt> point at 0 0
\setplotarea x from -6.0 to 1.0, y from 2.0 to 8.0
\inboundscheckon
\linethickness 0.5pt

%% Frame
\axis bottom label {\boldmath$m_{3/2}$(GeV)}
      ticks in
            width <0.8pt> length <8.0pt>
            withvalues {$10^{-6}$} {$10^{-5}$} {$10^{-4}$} {$10^{-3}$} 
                       {$10^{-2}$} {$10^{-1}$} {$1$} {$10$} /
            at -6 -5 -4 -3 -2 -1 0 1 /
            width <0.5pt> length <4.0pt> at
     -5.699 -5.523 -5.398 -5.301 -5.222 -5.155 -5.097 -5.046
     -4.699 -4.523 -4.398 -4.301 -4.222 -4.155 -4.097 -4.046
     -3.699 -3.523 -3.398 -3.301 -3.222 -3.155 -3.097 -3.046
     -2.699 -2.523 -2.398 -2.301 -2.222 -2.155 -2.097 -2.046
     -1.699 -1.523 -1.398 -1.301 -1.222 -1.155 -1.097 -1.046
     -0.699 -0.523 -0.398 -0.301 -0.222 -0.155 -0.097 -0.046
     0.301 0.477 0.602 0.699 0.778 0.845 0.903 0.954
     /
/
\axis left label {\rotatebox[origin=c]{90}{\boldmath$T_R$(GeV)}}
      ticks in
            width <0.8pt> length <8.0pt>
            withvalues {$10^2$} {$10^3$} {$10^4$} {$10^5$} 
                       {$10^6$} {$10^7$} {$10^8$} /
            at 2 3 4 5 6 7 8 /
            width <0.5pt> length <4.0pt> at
     2.301 2.477 2.602 2.699 2.778 2.845 2.903 2.954
     3.301 3.477 3.602 3.699 3.778 3.845 3.903 3.954
     4.301 4.477 4.602 4.699 4.778 4.845 4.903 4.954
     5.301 5.477 5.602 5.699 5.778 5.845 5.903 5.954
     6.301 6.477 6.602 6.699 6.778 6.845 6.903 6.954
     7.301 7.477 7.602 7.699 7.778 7.845 7.903 7.954
     /
/
\axis top
      ticks in
            width <0.8pt> length <8.0pt>
            at -6 -5 -4 -3 -2 -1 0 1 /
            width <0.5pt> length <4.0pt> at
     -5.699 -5.523 -5.398 -5.301 -5.222 -5.155 -5.097 -5.046
     -4.699 -4.523 -4.398 -4.301 -4.222 -4.155 -4.097 -4.046
     -3.699 -3.523 -3.398 -3.301 -3.222 -3.155 -3.097 -3.046
     -2.699 -2.523 -2.398 -2.301 -2.222 -2.155 -2.097 -2.046
     -1.699 -1.523 -1.398 -1.301 -1.222 -1.155 -1.097 -1.046
     -0.699 -0.523 -0.398 -0.301 -0.222 -0.155 -0.097 -0.046
     0.301 0.477 0.602 0.699 0.778 0.845 0.903 0.954
     /
/
\axis right label { }
      ticks in
            width <0.8pt> length <8.0pt>
            at 2 3 4 5 6 7 8 /
            width <0.5pt> length <4.0pt> at
     2.301 2.477 2.602 2.699 2.778 2.845 2.903 2.954
     3.301 3.477 3.602 3.699 3.778 3.845 3.903 3.954
     4.301 4.477 4.602 4.699 4.778 4.845 4.903 4.954
     5.301 5.477 5.602 5.699 5.778 5.845 5.903 5.954
     6.301 6.477 6.602 6.699 6.778 6.845 6.903 6.954
     7.301 7.477 7.602 7.699 7.778 7.845 7.903 7.954
     /
/

\put {(b)} at -5 7

%% Data Plot
\setplotsymbol (.)
\setquadratic

%% m_\lambda=1000GeV, M_X=10^5GeV
\setsolid
\plot
	-5.76474        2 
	-5.65325        2.05 
	-5.51905        2.1 
	-5.37059        2.15 
	-5.21897        2.2 
	-5.07345        2.25 
	-4.93977        2.3 
	-4.82059        2.35 
	-4.71652        2.4 
	-4.62697        2.45 
	-4.5507        2.5 
	-4.48622        2.55 
	-4.43196        2.6 
	-4.38636        2.65 
	-4.34797        2.7 
	-4.31549        2.75 
	-4.28773        2.8 
	-4.26364        2.85 
	-4.24233        2.9 
	-4.22301        2.95 
	-4.20502        3. 
	-4.1878        3.05 
	-4.1709        3.1 
	-4.15392        3.15 
	-4.13656        3.2 
	-4.11856        3.25 
	-4.09971        3.3 
	-4.07986        3.35 
	-4.05888        3.4 
	-4.03667        3.45 
	-4.01316        3.5 
	-3.9883        3.55 
	-3.96193        3.6 
	-3.93282        3.65 
	-3.89233        3.7 
	-3.79907        3.75 
	-3.5554        3.8 
	-3.13394        3.85 
	-2.64993        3.9 
	-2.18954        3.95 
	-1.77813        4. 
	-1.41821        4.05 
	-1.1062        4.1 
	-0.837249        4.15 
	-0.60649        4.2 
	-0.409353        4.25 
	-0.241657        4.3 
	-0.0996047        4.35 
	0.0202352        4.4 
	0.120953        4.45 
	0.205318        4.5 
	0.275801        4.55 
	0.334584        4.6 
	0.383588        4.65 
	0.424477        4.7 
	0.458686        4.75 
	0.48743        4.8 
	0.511729        4.85 
	0.532429        4.9 
	0.550216        4.95 
	0.565647        5. 
	0.578974        5.05 
	0.590906        5.1 
	0.601542        5.15 
	0.611093        5.2 
	0.619723        5.25 
	0.627561        5.3 
	0.634705        5.35 
	0.641237        5.4 
	0.64722        5.45 
	0.652707        5.5 
	0.657744        5.55 
	0.662368        5.6 
	0.666612        5.65 
	0.670509        5.7 
	0.674083        5.75 
	0.677361        5.8 
	0.680366        5.85 
	0.683118        5.9 
	0.685637        5.95 
	0.687942        6. 
	0.69005        6.05 
	0.691976        6.1 
	0.693735        6.15 
	0.695341        6.2 
	0.696807        6.25 
	0.698143        6.3 
	0.699362        6.35 
	0.700472        6.4 
	0.701483        6.45 
	0.702404        6.5 
	0.703242        6.55 
	0.704005        6.6 
	0.704698        6.65 
	0.705328        6.7 
	0.705901        6.75 
	0.706422        6.8 
	0.706895        6.85 
	0.707324        6.9 
	0.707714        6.95 
	0.708068        7. 
	0.708389        7.05 
	0.70868        7.1 
	0.708943        7.15 
	0.709183        7.2 
	0.709399        7.25 
	0.709596        7.3 
	0.709774        7.35 
	0.709935        7.4 
	0.710081        7.45 
	0.710213        7.5 
	0.710333        7.55 
	0.710441        7.6 
	0.710539        7.65 
	0.710627        7.7 
	0.710707        7.75 
	0.710779        7.8 
	0.710845        7.85 
	0.710904        7.9 
	0.710957        7.95 
	0.711006        8. 
/

%% m_\lambda=1000GeV, M_X=10^6GeV
\setdashes
\plot
	-5.76474        2 
	-5.65325        2.05 
	-5.51905        2.1 
	-5.37059        2.15 
	-5.21897        2.2 
	-5.07345        2.25 
	-4.93977        2.3 
	-4.82059        2.35 
	-4.71652        2.4 
	-4.62697        2.45 
	-4.5507        2.5 
	-4.48622        2.55 
	-4.43196        2.6 
	-4.38636        2.65 
	-4.34797        2.7 
	-4.31549        2.75 
	-4.28773        2.8 
	-4.26364        2.85 
	-4.24233        2.9 
	-4.22301        2.95 
	-4.20502        3. 
	-4.1878        3.05 
	-4.1709        3.1 
	-4.15392        3.15 
	-4.13656        3.2 
	-4.11856        3.25 
	-4.09971        3.3 
	-4.07986        3.35 
	-4.05888        3.4 
	-4.03667        3.45 
	-4.01316        3.5 
	-3.98831        3.55 
	-3.96208        3.6 
	-3.93446        3.65 
	-3.90546        3.7 
	-3.87509        3.75 
	-3.84338        3.8 
	-3.81037        3.85 
	-3.7761        3.9 
	-3.74061        3.95 
	-3.70398        4. 
	-3.66625        4.05 
	-3.62748        4.1 
	-3.58775        4.15 
	-3.54711        4.2 
	-3.50562        4.25 
	-3.46335        4.3 
	-3.42036        4.35 
	-3.37671        4.4 
	-3.33244        4.45 
	-3.28761        4.5 
	-3.24228        4.55 
	-3.19648        4.6 
	-3.15023        4.65 
	-3.10343        4.7 
	-3.05537        4.75 
	-3.00306        4.8 
	-2.93728        4.85 
	-2.83767        4.9 
	-2.67755        4.95 
	-2.4484        5. 
	-2.17349        5.05 
	-1.8856        5.1 
	-1.60626        5.15 
	-1.34489        5.2 
	-1.10413        5.25 
	-0.883726        5.3 
	-0.682419        5.35 
	-0.498739        5.4 
	-0.331276        5.45 
	-0.178753        5.5 
	-0.0400077        5.55 
	0.0860261        5.6 
	0.200347        5.65 
	0.303898        5.7 
	0.397573        5.75 
	0.482221        5.8 
	0.558639        5.85 
	0.627576        5.9 
	0.689728        5.95 
	0.745741        6. 
	0.795061        6.05 
	0.840514        6.1 
	0.881449        6.15 
	0.918317        6.2 
	0.951527        6.25 
	0.981448        6.3 
	1.00841        6.35 
	1.03272        6.4 
	1.05465        6.45 
	1.07443        6.5 
	1.09228        6.55 
	1.10839        6.6 
	1.12295        6.65 
	1.13611        6.7 
	1.148        6.75 
	1.15875        6.8 
	1.16848        6.85 
	1.17728        6.9 
	1.18525        6.95 
	1.19246        7. 
	1.199        7.05 
	1.20492        7.1 
	1.21028        7.15 
	1.21514        7.2 
	1.21955        7.25 
	1.22354        7.3 
	1.22717        7.35 
	1.23045        7.4 
	1.23342        7.45 
	1.23612        7.5 
	1.23857        7.55 
	1.24079        7.6 
	1.2428        7.65 
	1.24462        7.7 
	1.24628        7.75 
	1.24777        7.8 
	1.24913        7.85 
	1.25036        7.9 
	1.25148        7.95 
	1.25249        8. 
/

%% m_\lambda=1000GeV, M_X=10^7GeV
\setdots <2pt>
\plot
	-5.76474        2 
	-5.65325        2.05 
	-5.51905        2.1 
	-5.37059        2.15 
	-5.21897        2.2 
	-5.07345        2.25 
	-4.93977        2.3 
	-4.82059        2.35 
	-4.71652        2.4 
	-4.62697        2.45 
	-4.5507        2.5 
	-4.48622        2.55 
	-4.43196        2.6 
	-4.38636        2.65 
	-4.34797        2.7 
	-4.31549        2.75 
	-4.28773        2.8 
	-4.26364        2.85 
	-4.24233        2.9 
	-4.22301        2.95 
	-4.20502        3. 
	-4.1878        3.05 
	-4.1709        3.1 
	-4.15392        3.15 
	-4.13656        3.2 
	-4.11856        3.25 
	-4.09971        3.3 
	-4.07986        3.35 
	-4.05888        3.4 
	-4.03667        3.45 
	-4.01316        3.5 
	-3.98831        3.55 
	-3.96208        3.6 
	-3.93446        3.65 
	-3.90546        3.7 
	-3.87509        3.75 
	-3.84338        3.8 
	-3.81037        3.85 
	-3.7761        3.9 
	-3.74061        3.95 
	-3.70398        4. 
	-3.66625        4.05 
	-3.62748        4.1 
	-3.58775        4.15 
	-3.54711        4.2 
	-3.50562        4.25 
	-3.46335        4.3 
	-3.42036        4.35 
	-3.37671        4.4 
	-3.33244        4.45 
	-3.28761        4.5 
	-3.24228        4.55 
	-3.19648        4.6 
	-3.15026        4.65 
	-3.10365        4.7 
	-3.0567        4.75 
	-3.00943        4.8 
	-2.96189        4.85 
	-2.91408        4.9 
	-2.86605        4.95 
	-2.81781        5. 
	-2.76939        5.05 
	-2.7208        5.1 
	-2.67206        5.15 
	-2.62318        5.2 
	-2.57419        5.25 
	-2.52509        5.3 
	-2.47589        5.35 
	-2.4266        5.4 
	-2.37724        5.45 
	-2.32781        5.5 
	-2.27832        5.55 
	-2.22878        5.6 
	-2.17918        5.65 
	-2.12953        5.7 
	-2.07978        5.75 
	-2.02964        5.8 
	-1.97781        5.85 
	-1.9198        5.9 
	-1.84322        5.95 
	-1.72375        6. 
	-1.53535        6.05 
	-1.27801        6.1 
	-0.982146        6.15 
	-0.680655        6.2 
	-0.392932        6.25 
	-0.127026        6.3 
	0.114946        6.35 
	0.333464        6.4 
	0.530033        6.45 
	0.706493        6.5 
	0.864725        6.55 
	1.00653        6.6 
	1.13357        6.65 
	1.24738        6.7 
	1.34934        6.75 
	1.44068        6.8 
	1.52253        6.85 
	1.5959        6.9 
	1.66168        6.95 
	1.72068        7. 
	1.77242        7.05 
	1.81993        7.1 
	1.86259        7.15 
	1.90091        7.2 
	1.93534        7.25 
	1.9663        7.3 
	1.99416        7.35 
	2.01923        7.4 
	2.0418        7.45 
	2.06215        7.5 
	2.08049        7.55 
	2.09704        7.6 
	2.11197        7.65 
	2.12545        7.7 
	2.13763        7.75 
	2.14864        7.8 
	2.15859        7.85 
	2.1676        7.9 
	2.17574        7.95 
	2.18312        8. 
/

\endpicture$$
\medskip
\caption{The upper bound on the reheat temperature $T_R$ from
$\Omega_{3/2}h^2<1$ as a function of the gravitino mass $m_{3/2}$
in the gauge mediated models. We used the MSSM gaugino mass
$m_\lambda=300$ GeV in (a) and $m_\lambda=1$TeV in (b),
and the messenger mass $M_X=10^5$ GeV (solid line),
$M_X=10^6$ GeV (dashed line), $M_X=10^7$ GeV (dotted line), respectively.
}
\end{figure}


\begin{references}

\def\jhep#1#2#3{JHEP {#1}, #2 (#3)}
\def\npb#1#2#3{Nucl.\ Phys.\ {B#1}, #2 (#3)}
\def\prd#1#2#3{Phys.\ Rev.\ {D#1}, #2 (#3)}
\def\prl#1#2#3{Phys.\ Rev.\ Lett.\ {#1}, #2 (#3)}
\def\plb#1#2#3{Phys.\ Lett.\ {B#1}, #2 (#3)}
\def\prt#1#2#3{Phys.\ Rep.\ {#1}, #2 (#3)}
\def\zpc#1#2#3{Z.\ Phys.\ {C#1}, #2 (#3)}

\bibitem{Nilles}
For a review, see H. P. Nilles, \prt{110}{1}{1984}.
% Supersymmetry, supergravity and particle physics

\bibitem{Giudice-Rattazzi}
For a review, see G. F. Giudice and R. Rattazzi, hep-ph/9801271
% Theories with gauge-mediated supersymmetry breaking

\bibitem{Fayet}
P. Fayet, \plb{175}{471}{1986}.

\bibitem{Moroi-Murayama-Yamaguchi}
T. Moroi, H. Murayama and M. Yamaguchi, \plb{303}{289}{1993}.

\bibitem{Gouvea-Moroi-Murayama}
A. de Gouvi\^ea, T. Moroi and H. Murayama, \prd{56}{1281}{1997}.
% Cosmology of supersymmetric models with low-energy gauge mediation

\bibitem{TLee} T. Lee, hep-ph/9812451.

\bibitem{Dine-Nelson}
M. Dine and A. E. Nelson, \prd{48}{1277}{1993};
% Dynamical supersymmetry breaking at low-energies
M. Dine, A. E. Nelson and Y. Shirman, \prd{51}{1362}{1995};
% Low-energy dynamical supersymmetry breaking simplified
M. Dine, A. E. Nelson, Y. Nir and Y. Shirman, \prd{53}{2658}{1996}.
% New tools for low-energy dynamical supersymmetry breaking

\bibitem{others1} 
K. Choi and K. Y. Lee, \prd{54}{6591}{1996};
E. Poppitz and S. P. Trivedi, \prd{55}{5508}{1997};
N. Arkani-Hamed, J. March-Russell and H. Murayama, \npb{509}{3}{1998}.

\bibitem{others2}
K. I. Izawa and T. Yanagida,  Prog. Theor. Phys.  95, 829 (1996);
K. Intriligator and S. Thomas, \npb{473}{121}{1996}



\bibitem{Lee-Wu}
T. Lee and G.-H. Wu, \plb{447}{83}{1999}.
% Nonlinear supersymmetric effective lagrangian and goldstino interactions at
% high evenrgies

\bibitem{Brignole-Ferruccio-Zwirner}
A. Brignole, F. Feruglio and F. Zwirner, \jhep{9711}{001}{1997}.
% On the effective interactions of a light gravitino with matter fermions

\bibitem{pagels}
H. Pagels and J. R. Primack, \prl{48}{223}{1982}

\bibitem{ellis}
J. Ellis, J. E. Kim and D. V. Nanopoulos,
\plb{145}{181}{1984}; J. Ellis, G. B. Gelmini, J. L. Lopez, D. V. Nanopoulos
and S. Sarkar, \npb{373}{399}{1992}.

% \bibitem{Izawa-Nomura-Yanagida}
% Izawa K.-I., Y. Nomura and T. Yanagida, hep-ph/9901345.
% A gauge-mediation model of dynamical SUSY breaking
% with a wide range of the gravitino mass

\end{references}
\end{document}